\documentclass[preprint]{aastex}
\usepackage{emulateapj5}
\submitted{Astrophysical Journal Letters, in press}

\def\msun{{\rm\, M}_\odot}
\def\spose#1{\hbox to 0pt{#1\hss}}\def\lta{\mathrel{\spose{\lower 3pt\hbox{$\mathchar"218$}}
     \raise 2.0pt\hbox{$\mathchar"13C$}}}
\def\gta{\mathrel{\spose{\lower 3pt\hbox{$\mathchar"218$}}
     \raise 2.0pt\hbox{$\mathchar"13E$}}}
\def\kms{\,{\rm km\,s}^{-1}}
\def\etal{et al.\ }

\font\fivebmi=cmmib6
\font\sixbmi=cmmib6     \skewchar\sixbmi='177
\font\ninebmi=cmmib10 at 9pt    \skewchar\ninebmi='177
\newfam\bmifam
\textfont\bmifam=\ninebmi
\scriptfont\bmifam=\sixbmi
\scriptscriptfont\bmifam=\fivebmi

\mathchardef\alpha="710B
\mathchardef\beta="710C
\mathchardef\gamma="710D
\mathchardef\delta="710E
\mathchardef\epsilon="710F
\mathchardef\zeta="7110
\mathchardef\eta="7111
\mathchardef\theta="7112
\mathchardef\iota="7113
\mathchardef\kappa="7114
\mathchardef\lambda="7115
\mathchardef\mu="7116
\mathchardef\nu="7117
\mathchardef\xi="7118
\mathchardef\pi="7119
\mathchardef\rho="711A
\mathchardef\sigma="711B
\mathchardef\tau="711C
\mathchardef\upsilon="711D
\mathchardef\phi="711E
\mathchardef\chi="711F
\mathchardef\psi="7120
\mathchardef\omega="7121
\mathchardef\varepsilon="7122
\mathchardef\vartheta="7123
\mathchardef\varpi="7124
\mathchardef\varrho="7125
\mathchardef\varsigma="7126
\mathchardef\varphi="7127

\def\chaphead{}
\newcount\eqnumber
\def\today{\ifcase\month\or
 January\or February\or March\or April\or May\or June\or
 July\or August\or September\or October\or November\or December\fi
 \space\number\day, \number\year}

\def\eqnam#1{\xdef#1{(\chaphead\the\eqnumber}}

\def\newe{(\hbox{\chaphead\the\eqnumber})\global\advance\eqnumber by 1}
\def\firste{(\hbox{\chaphead\the\eqnumber a})\global\advance\eqnumber by 1}
\def\laste#1{\advance\eqnumber by -1%
        (\hbox{\chaphead\the\eqnumber #1})\advance\eqnumber by 1}

\def\refe#1{\advance\eqnumber by -#1 (\chaphead\the\eqnumber
     \advance\eqnumber by #1 }

\def\i{\relax\ifmmode{\rm i}\else\char16\fi}

\def\frac#1#2{{\textstyle{#1\over#2}}}

\def\d{{\rm d}}
\def\dddot#1{\ddot#1\kern-1.4pt\dot{\phantom{#1}}\kern-3pt}

\def\spose#1{\hbox to 0pt{#1\hss}}

\def\=#1{\overline{#1}}

\def\lta{\mathrel{\spose{\lower 3pt\hbox{$\mathchar"218$}}
     \raise 2.0pt\hbox{$\mathchar"13C$}}}
\def\gta{\mathrel{\spose{\lower 3pt\hbox{$\mathchar"218$}}
     \raise 2.0pt\hbox{$\mathchar"13E$}}}

\def\kms{{\rm\,km\,s^{-1}}}

\def\msun{{\rm\,M_\odot}}

\def\pc{{\rm\,pc}}
\def\cm{{\rm\,cm}}
\def\yr{{\rm\,yr}}

\def\myr{{\rm\,Myr}}

\def\annrev #1 #2 {ARA\&A, #1, #2}
\def\aa #1 #2 {A\&A, #1, #2}
\def\aasupp #1 #2 {A\&AS, #1, #2}
\def\aj #1 #2 {AJ, #1, #2}
\def\apj #1 #2 {ApJ, #1, #2}
\def\apjlett #1 #2 {ApJ, #1, #2}
\def\apjsupp #1 #2 {ApJS, #1, #2}
\def\ban #1 #2 {Bull.\ Astron.\ Inst.\ Netherlands, #1, #2}
\def\mn #1 #2 {MNRAS, #1, #2}
\def\nature #1 #2 {Nat, #1, #2}
\def\pasj #1 #2 {PASJ, #1, #2}
\def\pasp #1 #2 {PASP, #1, #2}

\shorttitle{Origin of Galactic Center HeI Star Cluster}
\shortauthors{Ortwin Gerhard}

\begin{document}
\title{The Galactic Center He I Stars: \\
       Remains of a Dissolved Young Cluster?}
\author{Ortwin Gerhard}
\affil{Astronomical Institute of the University of Basel, Venusstrasse
7, CH-4102 Binningen, Switzerland}


\begin{abstract}
A massive young star cluster, initially embedded in its parent
molecular cloud, will spiral into the Galactic Center from $\lta
30m_6^{1/2}\pc$ during the life-time of its most massive stars, if
the combined total mass is $\sim 10^6m_6\msun$. On its way inwards the
system loses most of its mass to the strong tidal field, until the
dense cluster core of high-mass stars is finally disrupted near the
central black hole.  A simple model is presented to argue that this
scenario may under plausible conditions explain the observed location
and rotation of the Galactic Center HeI stars. Accretion of star
clusters into the Galactic Center could be recurrent, and play an
important role in regulating the activity of Sgr A$^\ast$.
\end{abstract}


\section{Introduction}

The central parsec of the Galaxy contains a cluster of young stars,
including some 15 very luminous HeI emission line stars (Krabbe \etal
1995), as well as many less massive O and probably B stars (Eckart
\etal 1999). The HeI stars are believed to be post-main sequence
supergiant stars of $\sim 20-100\msun$, close to the Wolf-Rayet stage
(Najarro \etal 1997). The most massive of these stars have a total age
of $\lta 3 \myr$, while the less luminous stars could have ages up to
$\sim 8\myr$. Krabbe \etal (1995) have argued that the most likely
origin of these stars is in a small starburst $\sim 3-7\myr$ ago, in
which $\gta 10^4\msun$ of young stars were created.

In situ formation of these stars is problematic, however, because of
the strong tidal field of the Galactic nuclear bulge and central black hole.
The tidal forces from the
Galactic nucleus alone are sufficient to unbind gas clouds with
densities $n_{\rm crit}<10^7\cm^{-3} (1.6\pc/R_G)^{1.8}$ at
galactocentric radii $R_G$ (e.g., Morris 1993), while clouds in the
nuclear gas disk have densities of $10^4 - 3\times 10^5 \cm^{-3}$
(Genzel 1989, G\"usten 1989).  Hence Morris (1993) has argued that the
formation of the central star cluster must have been externally
triggered to achieve the required high densities, perhaps in a cloud
collision. Two clouds coming in from $\sim 10\pc$ would need to
collide near the gravitational radius of the black hole ($\sim
0.5\pc$), at a velocity of several hundreds of $\kms$; it is unclear
whether the required densities and a high efficiency of star formation
could be reached. Another model in which the massive stars
form through collisions and mergers of lower mass stars in the
high-density nuclear cluster now appears unlikely, both because too
few massive stars would form (Lee 1994), and because similar stars
have been found in the Arches and Quintuplet clusters some $30\pc$
away from the center (Nagata \etal 1995, Cotera \etal 1996, Figer
\etal 1999a).

This letter therefore explores an alternative idea: that the young
stars now seen in the Galactic Center (GC) formed
further out in a massive star cluster that subsequently spiralled
into the nucleus and tidally dissolved. One of the exciting results
from HST has been the discovery of young star clusters in a variety of
starburst environments (e.g., Whitmore \& Schweizer 1995, O'Connell
\etal 1995, Oestlin \etal 1998). We also know now that nuclear star
clusters are common in spiral galaxies (Carollo \etal 1998, Matthews
\etal 1999).  The Arches and Quintuplet clusters at $\sim 30\pc$
distance from the GC testify that a similar star formation mode has
been occurring in the nuclear disk of the Galaxy.  Both clusters have
ages of a few megayears and estimated total masses (extrapolating the
IMF to $1\msun$) of $\sim 10^4 \msun$ (Figer \etal 1999b). The orbits
of massive clusters evolve by dynamical friction against field stars
in sufficiently dense stellar systems (Tremaine, Ostriker \& Spitzer
1975). Here I show that a massive cluster formed well outside the
central few parsecs will indeed spiral into the center within the lifetime
of its most massive stars, losing much of its mass on the way inwards
until finally its dense core is disrupted deep in the nucleus.

\section{Massive Cluster Infall}

Suppose a massive star cluster is formed at initial galactocentric
radius $R_i=30 R_{30}\pc$. The mass distribution of the nuclear
bulge in $2\pc<R_G<30\pc$ can be approximated as an isothermal sphere
(Genzel, Hollenbach \& Townes 1994, Fig.~7.1), with circular velocity
$v_c=130 v_{130} \kms$ in this range of radii, and
\begin{equation}\label{eqmbulge}
  M_G(R_G)=3.9\times 10^6 v_{130}^2 R_G[\pc ] \msun.
\end{equation}
Eq.~(\ref{eqmbulge}) implies that the mean density within
$10-30\pc$ corresponds to $10^{5.5}-10^{4.5}$ H-atoms per $\cm^3$, so
in this region molecular clouds with the densities observed in the GC
may indeed collapse and form stars.

The newly formed cluster will lose orbital energy by dynamical
friction against the nuclear bulge stars and spiral into the center
on a time-scale 
\begin{equation}\label{eqtfric}
\tau_{\rm df}={1.17\over \ln(0.4N)}{R_i^2v_c\over Gm_c}
        =3.1 \times 10^6 R_{30}^2 v_{130} m_6^{-1} \lambda_{10}^{-1} \yr,
\end{equation}
where the cluster mass is $m_c=10^6 m_6 \msun$ and 
$\lambda_{10} =\ln(0.4N)/10 > 1$ with $N$ the effective number of 
field stars (Binney \& Tremaine 1987). Thus a massive
cluster, $m_c \sim 10^6\msun$, will indeed spiral into the center from
$R_i\lta 30\pc$ within the lifetime of its most massive stars.
$\tau_{\rm df}$ corresponds to only a few initial rotation times
around the center; at $R_i$
\begin{equation}\label{eqtrot}
\tau_{\rm rot}\equiv 2\pi R_i/v_c = 1.4\times 10^6 R_{30} v_{130}^{-1} \yr.
\end{equation}

The friction time-scale $\tau_{\rm df}$ is dominated by the time spent
at large radii. In these initial phases the cluster will be embedded
within its parent molecular cloud while both are dragged inwards
together.  Because the part of the molecular cloud not converted into
stars is conceivably the major part of the total initial mass, the
dynamical friction on the cloud (Stark \etal 1991) contributes to the
orbital evolution of the cluster.  $\tau_{\rm df}$ is shorter or
comparable to the time-scales of other processes that might
separate the surrounding gas cloud from the cluster. Magnetic
friction (Elmegreen 1981) brakes the gas cloud in $\sim 3\times
10^7\yr$ (Morris 1996), assuming a mG field continuously permeates the
cloud and rotates more slowly than the cloud by a fraction of the
circular velocity. Mass loss from bulge stars (Jenkins \& Binney
1994), even if it can cool and settle into the disk plane, amounts to
only $\sim 5\times 10^4\msun$ over $3\times 10^6\yr$ into the central
$30\pc$, so dilutes the cloud angular momentum only slightly.  Most
dangerous for the cloud is the radiation from the massive stars, which
drives an expanding HII region; e.g., for the $6\times
10^{51}$ Ly c photons per second from a $3\times 10^5\msun$ cluster
(Efstathiou \etal 2000), and a cloud of $1.7\times 10^6\msun$ with
density $10^{4.5} \cm^{-3}$, the HII region front will reach the
surface after $2.5\times 10^6\yr$ in standard spherical
theory (Spitzer 1978).  Thereafter the gas expands and is tidally stripped,
but will take a further $\gta$ 1-2 orbital times before it also
gravitationally decouples from the cluster. To separate
most of the cloud from the cluster will take longer if the cluster is
located near the cloud surface or if the cloud is fractal-like.
Thus the surrounding molecular cloud could contribute to
the dynamical friction for $\sim \tau_{\rm df}$, certainly in the crucial
initial phases, and is likely to be eventually left behind. In detail
this depends on the parameters of the cloud and cluster, but
the most massive clouds are the most robust.

As it spirals to the Galactic center, the parent cloud and embedded
cluster will constantly lose mass because of the strong tidal
field. We can estimate the effect of this mass loss on the friction
time with a simple model in which the initial mass $m_{ci}$ of the
combined cloud and cluster is distributed according to an isothermal
profile, tidally limited at radius $r=r_{ti}$:
\begin{equation}
  m_c(r)=m_{ci}(r/r_{ti}),
\end{equation}
\begin{equation}\label{eqrinitial}
  r_{ti}=\left( {m_{ci}\over M_G(R_i)} \right)^{1/3} \!\! R_i
        = 6.2\, m_6^{1/3} v_{130}^{-2/3} R_{30}^{2/3} \pc.
\end{equation}
For comparison, the half-mass radius of the present Arches cluster
($\sim 10^4\msun$) is $\sim 0.2\pc$ (Figer \etal 1999a).  As
the cluster spirals in, successive outer shells of the mass
distribution are peeled off by the tidal field. The division between
cluster and cloud need not be specified now, but it is clear
that at some stage the cloud envelope will have been removed and the
cluster proper will begin to be stripped. If $R_G$ is the current
galactocentric radius, the tidal radius decreases according to
\begin{equation}\label{eqrtidal}
  r_t/R_G = r_{ti}/R_i,
\end{equation}
and if internal evolution is neglected, the cluster mass decreases as
\begin{equation}\label{eqclmass}
  m_c(R_G) = (R_G/R_i) \, m_{ci}.
\end{equation}
When the entire cluster is finally disrupted at radius $R_{\rm dis}$,
only a fraction $\sim R_{\rm dis}/R_i$ of the initial mass $m_{ci}$
will have arrived at $R_{\rm dis}$.

We can now write a modified dynamical friction equation using the same
arguments as in Binney \& Tremaine (1987), and assuming 
that the instantaneous specific angular momentum loss
due to the frictional force is the same for stripped material and
material that remains bound to the cluster. This gives
$R_G \d{R_G}/\d{t} =-0.428 G\ln(0.4N) v_c^{-1} m_{ci} (R_G/R_i)$,
the solution of which is
\begin{eqnarray}\label{eqtfricprimed}
\tau'_{\rm df}(R_i,R_G)= R_i R_G v_c/ 0.428\,\ln(0.4N)\,G m_{ci}
        & \\
  = 6.2 \times 10^6 (R_G/R_i) R_{30}^2 v_{130} m_6^{-1} \lambda_{10}^{-1}
                                \yr.& \nonumber
\end{eqnarray}
Thus in this simple model the total time to spiral in from radius
$R_i$ is just twice that when the mass $m_{ci}$ remains constant
[eq.~(\ref{eqtfric})], and the time taken from radius $R_G<R_i$ for a
cluster that started out tidally limited at $R_i$ is linearly
proportional to $R_G$. To reach the center in $\sim 3\times 10^6\yr$
from $R_i\simeq 30\pc$ ($10\pc$), a mass-losing cloud-cluster system
must have an initial mass $m_{ci}\simeq 2\times 10^6\msun$ ($\simeq
2\times 10^5\msun$). The core of a cluster with initial mass $2\times
10^6\msun$ formed at $R_i\simeq 10\pc$ would have spiralled into the
center after 300,000 years.
These time-scales may be slightly overestimated because we have
neglected torques from previously stripped material and {\it its}
wake, left behind on orbits of lower frequency, as well as any
additional drag from the nuclear gas disk.  Perturbed material within
a factor 1.5-2 in galactocentric radius contributes to the frictional 
drag (Tremaine \& Weinberg 1984); over this radial range the cluster 
loses about half of its mass [eq.~(\ref{eqclmass})].

One constraint is that the cluster must not evaporate in the strong
tidal field of the nuclear bulge before it reaches the center.  In the
$m_c=2\times 10^4\msun$ evolutionary models for the Arches and
Quintuplet clusters by Kim \etal (1999) the evaporation time is
several $\myr$.  Evaporation occurs on a time-scale proportional to
the half-mass relaxation time 
\begin{equation}
\label{eqtrhalf} t_{rh}
       = \, 2.0 \times 10^8 \, m_{6}^{1/2} \, r_{1}^{3/2} \,
         m_{\ast,\odot}^{-1}\, \lambda_{10}^{-1} \, \yr,
\end{equation}
(Spitzer \& Hart 1971) but depends also strongly on the strength of
the tidal field. Here $r_h=r_1\pc$ is the cluster half-mass radius and
$m_{\ast}=m_{\ast,\odot}\msun$ the average stellar mass.  $t_{rh}$ is
sensitive to the uncertain half-mass radius.  For the Arches cluster
Figer \etal (1999a) estimated $r_h=0.2\pc$ from the observed high-mass
stars, i.e., without correction for initial or dynamical mass
segregation. We can rewrite $t_{rh} \propto m_c^{1/2}r_h^{3/2}\propto
m_c\rho^{-1/2}$ and consider the mean density to be fixed by the
external tidal field. This suggests independently that only clusters
significantly more massive or formed at smaller radii than the Arches
cluster can reach the central parsec.

\section{Final Tidal Disruption and the He I Star Cluster}

The preceding discussion shows that only the inner parts of a massive
young star cluster will spiral into the Galactic center proper, while
its outer parts and the cloud envelope will be left behind and
distributed further out.  There are good reasons to believe that the
massive stars will be concentrated towards the cluster core. There is
some evidence that high-mass stars form predominantly in or near the
cores of young clusters (Fischer \etal 1998, Bonnell \& Davies 1998).
After star formation is complete, dynamical mass segregation will
further accentuate the central concentration of massive stars, acting
on their two-body relaxation time-scale. Thus the most massive cluster
stars end up in the GC, as is needed to explain
the observed distribution of HeI stars.

Under what conditions is the density of the cluster core
high enough to reach the central parsec?  The GC HeI stars are
observed at galactocentric radii $R_G=1-10''= 0.04-0.4\pc$ and their
proper motions and radial velocities show a coherent rotation pattern
(Genzel \etal 2000). In the present scenario this would be interpreted
as tracing the last orbit of the cluster core before final disruption,
giving a radius of disruption $R_{\rm dis}\lta 0.4\pc$.  In this
radial range the tidal force of the central black hole dominates that
from the nuclear bulge. We can thus estimate the required mean density
of the cluster core at $R_{\rm dis}$ as
\begin{eqnarray}
  \rho_{\rm dis}={3m_c(R_{\rm dis})/4\pi r_t^3(R_{\rm dis})}
                = {6M_\bullet/4\pi R_{\rm dis}^3}& \\
         \qquad = 2.2\times 10^7 M_3
        \left(R_{\rm dis}/0.4\pc\right)^{-3} \msun \pc^{-3},& \nonumber
\end{eqnarray}
where we have written the black hole mass as $M_\bullet = 3\times 10^6
M_3 \msun$ (Genzel \etal 2000). Notice the sensitivity to the
disruption radius.

The observed average stellar density in the Arches cluster,
extrapolated to include stars down to $1\msun$, is $6.3\times
10^5\msun/\pc^3$ (Figer \etal 1999a).  The $2\times 10^4\msun$ models
of Kim \etal (1999) show that dynamical evolution and mass
segregation lead to the formation of a core of high mass stars with
density reaching $\rho_c \simeq 10^7\msun/\pc^3$ in the mild core
collapse stage, after $\sim 1\myr$ evolution.  Central densities of
$\sim 10^7\msun \pc^{-3}$ can thus be reached in later evolutionary
stages, high enough for the core to survive to $R_{\rm dis}\sim
0.4\pc$, but significantly larger densities (required for smaller
$R_{\rm dis}$) would seem problematic.

The time-scale for this evolution is a small number of relaxation
times for the massive stars (Chernoff \& Weinberg 1990).  If evolution
is too fast, the core may reexpand before the cluster has time
to reach the GC; this would be the case for the Arches
cluster models.  For clusters with initial masses $m_c\gta 3\times
10^5\msun$, the mass segregation time for $30\msun$ stars is $\gta
2\times 10^6\yr$, using (\ref{eqtrhalf}) and scaling radii such that
$r_{h}\propto m_c^{1/3}$.  Thus in the
initial stages dynamical evolution is relatively slow.  As the cluster
spirals in and loses mass, however, the evolution accelerates: the
half-mass relaxation time decreases because both the cluster mass and
half-mass radius decrease, and the external tidal field becomes
stronger, so that the mass evaporating per relaxation time
increases. Thus the most rapid dynamical evolution is expected to take
place when the stripped-down cluster approaches the center. At the
time of final disruption, its mass is a small fraction of the initial
mass, comparable to the model clusters studied by Kim \etal (1999) in
which the time-scale for core collapse is $\sim 10^6\yr$. Thus it is
not improbable that such a cluster reaches core collapse shortly
before it arrives at the Galactic Center. The high cluster core
density required for the core to reach $R_{\rm dis}\simeq 0.4 \pc$
would then most easily be maintained in a phase after core collapse
when the energy loss of the core from collisions is compensated by
the energy gain from three-body binaries. This requires the core to be
dominated by the 100 most massive stars (Binney \& Tremaine 1987),
previously concentrated into this volume by mass segregation.

After the high-density cluster core has reached $R_{\rm dis}\sim
0.4\pc$ and begins to disintegrate, it continues to spiral inwards
until torques from its friction wake and the material lost previously
(see \S2) become ineffective, that is, until the debris has spread in
angle by $\Delta\phi\sim\pi$. The time for the debris within $\alpha
r_t$ of the cluster center to spread by $\Delta\phi=\pi$ is
\begin{equation}
  t_{\rm spr}\equiv{\pi\over \left|\d\Omega\over\d R_G\right|_{R_{\rm dis}}
                                        \!\! 2\,\alpha\,r_t(R_{\rm dis})}
          = {t_{\rm rot}\over 6}\;{R_{\rm dis}\over \alpha\,r_t(R_{\rm dis})},
\end{equation}
which gives of order $t_{\rm rot}/\alpha$ according to
eqs.~(\ref{eqrinitial}), (\ref{eqrtidal}).  The rotation period near the
black hole is $t_{\rm rot}=13700 M_3^{-1/2} (R_G/0.4\pc)^{3/2} \yr$,
so for the cluster center $t_{\rm rot}/\alpha$ can be several
$10^4\yr$.  For comparison, the dynamical friction time
(\ref{eqtfric}) for $10^4\msun$ to spiral from $R_{\rm dis}=0.4\pc$
into the center completely is only $50000\yr$. This suggests that the
debris of the cluster core will spiral inwards significantly even
after the final disruption, and that therefore the rotation pattern
observed down to $R_G\simeq 0.15\pc$ is consistent with disruption at
radii $\sim 0.4\pc$. It will be interesting to simulate this
disruption event in the black hole's tidal field with numerical star
cluster models.

\section{Discussion}

In summary, the previous sections have shown that under plausible
conditions a young star cluster formed in the Galactic nuclear disk
may spiral into the GC within the life-time of its most massive stars,
and that the density in its core of high-mass stars can be such that
it would actually reach and dissolve in the region where the GC HeI
stars are observed. The important parameters are the initial
galactocentric radius, the cluster mass and radius, and the mass and
density of the surrounding molecular cloud. To reach the center from
initial radii $R_i\simeq 10-30\pc$, the cluster must be substantially
more massive than the Arches and Quintuplet clusters or the
frictional force must at least initially be augmented by the wake of
the surrounding molecular cloud, or both. Large cluster masses are
also required to avoid too rapid dynamical evaporation and evolution
in the strong tidal field.

A prediction of the present model for the origin of the GC HeI stars
is that the massive stars in the core of the disrupting young cluster
should end up at smaller galactocentric radii than lower mass stars in
the cluster envelope. Krabbe \etal (1995) estimate the total stellar
mass corresponding to the nuclear HeI star cluster as $\sim 1.5\times
10^4\msun$, assuming a mass function $N(m)\, dm \propto m^{-2} dm$
down to $m_\ast=1\msun$.  The number of low-mass stars associated with
the cluster is unknown, but a standard Salpeter IMF with the same
number of the highest mass stars, extrapolated to $m_\ast=0.1\msun$,
would give $\sim 10^5\msun$. Most of these stars should now be found
at galactocentric radii of a few parsec (some tens of arcsec) or
further out if the cluster was even more massive. Could these stars be
related to the young population inferred by Philipp \etal (1999), or
was the formation of the cluster just part of a larger-scale
starburst?

As in alternative models, the observed counterrotation of the HeI
stars with respect to Galactic rotation (Genzel \etal 2000) is not
inherent in the present scenario.  However, a large molecular cloud
entering the nuclear disk with orbital angular momentum roughly
antialigned with Galactic rotation would sooner or later suffer a
strong collision, and this in fact might be the event triggering the
formation of the cluster. The subsequent dynamical friction on the
cluster and its surrounding cloud envelope would then tend to
circularize the counterrotating orbit while the cluster spirals in,
explaining the rotating torus structure of the HeI stars observed now.
This observation would be more difficult to explain in models where
the GC young stars form in a cloud collision near their present
location, because some of these stars should then still reflect the
cloud's initial orbital motion.

How common could cluster accretion events be in the GC?  The presence
of $\sim 10^8\yr$ old AGB stars in the central parsec (Krabbe \etal
1995) suggests that it has happened before, and perhaps one of the
massive molecular clouds in the vicinity is the next candidate.
Clearly, a cluster accretion rate of $\sim 10^5\msun$ per $3\times
10^6\yr$ would build up substantial mass in the nuclear bulge over
time. The total mass inflow rate of $\sim 0.1\msun/\yr$ inferred from
dynamical friction on the GC giant molecular clouds on $x_2$-orbits at
$\sim 100\pc$ (Stark \etal 1991), and somewhat more from gas flow
through the ILR at $\sim 200\pc$ (Gerhard 1992, Morris \& Serabyn
1996) would supply sufficient gaseous material to maintain this rate
of cluster formation in addition to other distributed star formation.
Detailed observations of the young stellar population in the GC are
needed to see whether the process actually happens recurrently and
whether it is a substantial factor in building up the nuclear bulge.
Early-on, clusters would spiral in as far as there are field stars to
provide the background for dynamical friction. Later clusters would
come in further and build up a density gradient (assuming the angular
momentum can be further transported outwards), but only some would
reach the vicinity of the black hole. In the simple model
described in Section 2, where both the nuclear bulge and cluster
density profiles are $\propto r^{-2}$, the ingoing cluster's tidally
stripped material would also be distributed $\propto r^{-2}$.

Independent of the rate of cluster accretion, the recent arrival in the GC
of a cluster with a large number of massive stars would have profoundly
changed the physical conditions in the central parsec and may in fact
be responsible for the present lack of activity of Sgr A$^\ast$. If so,
stellar population studies of the frequency of cluster accretion in the
GC will have some general impact on understanding the nuclear activity
cycles in spiral galaxies.

\acknowledgements{It is a pleasure to thank J.~Gallagher, R.~Genzel,
and M.~Vietri for helpful information and discussion, and an anonymous
referee for his comments. This work was supported by grant 20-56888.99
from the Swiss National Science Foundation.}

\end{document}